\documentclass[fleqn,usenatbib]{mnras}

\usepackage{newtxtext,newtxmath}
\usepackage{xcolor}
\usepackage{mathtools}
\usepackage{amsmath}
\usepackage{braket}
\usepackage{graphicx}
\usepackage{subcaption}
\usepackage{hyperref}

\newcommand{\wigthreej}[6]{\left(\begin{array}{ccc} #1 & #2 &  #3 \\ #4 & #5 & #6 \end{array} \right) }

\DeclareRobustCommand{\orderof}{\ensuremath{\mathcal{O}}}

\def \ba{\begin{eqnarray}}
\def \ea{\end{eqnarray}}

\newcommand{\ourcodename}{\texttt{threej\_cosmo}}

\title[Fast computation of CMB coupling matrices]{Fast computation of temperature and polarization coupling matrices}

\author[G. Kiddier and S. Gratton]{
Georgia Kiddier,$^{1}$\thanks{E-mail: glk27@cam.ac.uk}
and Steven Gratton$^{1,2}$\thanks{E-mail: stg20@cam.ac.uk}
\\
$^{1}$Centre for Theoretical Cosmology, Department of Applied Mathematics and Theoretical Physics,\\
University of Cambridge, Wilberforce Road, Cambridge, CB3 0WA, UK\\
$^{2}$Kavli Institute for Cosmology Cambridge, Madingley Road, Cambridge, CB3 0HA, UK
}
\date{17 February 2026}

\pubyear{2026}
\begin{document}
\label{firstpage}
\pagerange{\pageref{firstpage}--\pageref{lastpage}}
\maketitle
\begin{abstract}
We present a fast and exact method for computing CMB mode--coupling matrices
based on an optimised evaluation of Wigner--$3j$ symbols.
The method exploits analytic structure in the relevant Wigner--$3j$ symbol configurations appearing in temperature and polarization coupling matrices, expressing all required quantities in terms of a small set of recurrence-generated values which are precomputed and stored in lookup tables. This approach reduces the computational cost of constructing the full coupling matrices whilst maintaining numerical accuracy.

We demonstrate the performance of the \ourcodename\space implementation using realistic survey masks from current CMB experiments. Relative to standard recursion-based approaches
used in existing pseudo-$C_\ell$ pipelines, the method achieves speedups of
$6$--$25\times$ in practical coupling-matrix constructions, with the largest gains occurring
at high multipoles. The algorithm admits efficient parallelisation on both CPUs and GPUs, the latter
providing additional acceleration, up to a further $\orderof(50)$ on modern hardware, without
altering the underlying formalism.

Beyond full matrix construction, the approach is naturally suited to applications in which only
a restricted set of $\ell_3$ modes is required for each $(\ell_1,\ell_2)$ pair, such as in the
computation of band-limited coupling matrices and analytic covariance terms. These features make
\ourcodename\space a practical backend for pseudo-$C_\ell$ estimation and related
calculations in next-generation CMB analysis pipelines.
\end{abstract}


\section{Introduction}

The cosmic microwave background (CMB), first discovered in 1965 \citep{penziasMeasurementExcessAntenna1965} and whose fluctuations were detected by the COBE satellite in 1992 \citep{smootStructureCOBEDifferential1992}, has become one of the most powerful probes of cosmological parameters. WMAP~\citep{hinshawNineyearWilkinsonMicrowave2013} and the Planck mission \citep{planckcollaborationPlanck2018Results2020} subsequently measured the CMB temperature and polarization anisotropies with unprecedented precision, constraining the standard $\Lambda$CDM model to percent-level accuracy. Much of the constraining power for cosmology from CMB anisotropies comes from their angular power spectra generically denoted $C_\ell$, with $\ell$ being angular multipole number. Forthcoming CMB experiments such as the Simons Observatory \citep{collaborationSimonsObservatoryScience2019} will provide an order of magnitude more data, reaching arcminute resolution across large sky areas. 

Observations of the CMB are contaminated by foreground emissions such as Galactic synchrotron, thermal dust, and unresolved point sources. To suppress these contaminants, standard analyses apply sky masks that disregard polluted regions of the sky. Such masking also induces mode coupling in spherical harmonic space.

The pseudo-$C_\ell$ method provides an efficient way to perform an analysis on the masked sky, relating the expectation of the observed pseudo-spectrum to the underlying power spectrum through a linear relationship that accounts for the masking of the sky~\citep{wandeltPseudoClMethodCosmic2001,hivonMASTERCMBAnisotropy2002,efstathiouMythsTruthsConcerning2004,alonsoUnifiedPseudoCellFramework2019}. 
This is done through the introduction of the mode-coupling matrix $K_{\ell_1 \ell_2}$, which can in turn be expressed in terms of the window function $W_\ell$, the harmonic power spectrum of the mask. Unlike maximum likelihood estimators, which scale with multipole as $\mathcal{O}\left(\ell_{\max}^6\right)$, the pseudo-$C_\ell$ method achieves a more tractable computational scaling of $\mathcal{O}\left(\ell_{\max}^3\right)$. This makes it well suited for high-resolution surveys such as Planck, ACT~\citep{naess2025atacamacosmologytelescopedr6}, and the Simons Observatory.

The MASTER algorithm \citep{hivonMASTERCMBAnisotropy2002} is a widely used implementation of this formalism and forms the basis for several public pipelines, including \texttt{pspy}\footnote{\url{https://github.com/simonsobs/pspy}}. Efficient and accurate evaluation of the coupling matrix $K_{\ell_1 \ell_2}$ is therefore important to the performance of pseudo-$C_\ell$-based estimators and is a key focus of this work. A significant amount of time is spent in the recursion formulae used to calculate Wigner--$3j$ symbols, see e.g.~\cite{edmondsAngularMomentumQuantum1957}, hereafter E57, that are contained within the elements of $K_{\ell_1\ell_2}$. At the heart of this computation is the algorithm of \citet{schultenExactRecursiveEvaluation1975}, which uses recurrence relations for exact evaluation of the $3j$ symbols of the form
\begin{equation}
\left( \begin{array}{ccc}
\ell_1 & \ell_2 & \ell_3 \\
m_1    & m_2    & m_3
\end{array} \right).
\label{eq:3j_general}
\end{equation}
This is implemented in the publicly available \texttt{SLATEC} Fortran library,
more specifically in the subroutine DRC3jj~\footnote{Available from
\url{http://www.netlib.org/slatec/src/drc3jj.f}}. For fixed $\ell_2$ and $\ell_3$ this routine evaluates the $3j$ symbol for all allowed values of $\ell_1$, with all other parameters fixed, satisfying $m_1+m_2+m_3=0$ and obeying the triangular inequality $\left| \ell_2-\ell_3 \right| \leq \ell_1 \leq \ell_2+\ell_3$. The subroutine is used for mode-coupling matrix calculation in current CMB pipelines, with $m_1=m_2=m_3=0$ for temperature and also $m_1=-2$, $m_2=2$, $m_3=0$ when including polarization. Pseudo-$C_\ell$ mode-coupling matrices of the same form as those used in CMB analyses also arise in projected LSS angular power spectra \citep{lizancosHarmonicAnalysisDiscrete2024}.

Wigner–$3j$ symbols also appear in the covariance of pseudo-$C_\ell$ estimates. While the exact expression is computationally expensive, involving sums over magnetic indices, useful approximations exist~\citep{efstathiouMythsTruthsConcerning2004,brownCMBTemperaturePolarisation2005}. Here the computation is reduced to one essentially equivalent to computing a mode coupling matrix. 

Some recent work has been done to speed up the computation of mode-coupling matrices, such as using the Toeplitz approximation for diagonal-constant matrices \citep{louisFastComputationAngular2020}. This approximation is shown to be highly accurate in power spectrum estimation, but is not exact for all multipoles in the coupling matrix because the structure is not exactly Toeplitz for all values of $\ell$.

Beyond pseudo-$C_\ell$ estimation, Wigner--$3j$ symbols also appear in several
other areas of CMB and large-scale structure analysis.
Lensing of the CMB induces off-diagonal covariance between
multipoles proportional to the lensing potential.

Wigner symbols with magnetic numbers  $(0,-1,1)$ then appear in the weight function of the optimal quadratic estimator for the lensing potential (see~\cite{okamotoCMBLensingReconstruction2003} and e.g.\
\cite{2014}). The same symbol also appears in the analysis of the aberration of the CMB (see e.g.~\cite{planckcollaborationPlanck2013Results2014}).

Related structures also appear in higher-order anaylses. In recent work on fast skew-spectrum and projected bispectrum estimators, the leading covariance contribution is shown to have a summation structure closely analogous to the pseudo-$C_\ell$ mode-coupling matrix~\citep{harscouetFastProjectedBispectra2025}.

These examples highlight that accelerating Wigner-symbol evaluation is 
beneficial not only for pseudo-$C_\ell$ power-spectrum estimation, but also more widely in cosmological analyses.
This paper outlines a method for simplified evaluation of such $3j$ symbols.
An associated code, \ourcodename, implements this approach\footnote{\url{https://github.com/gkiddier22/threej_cosmo}} and may be integrated into pseudo-$C_\ell$ estimation pipelines.

The paper is organized as follows. Section~\ref{sec:pseudo_cl} outlines the pseudo-$C_{\ell}$ estimation method and motivates the optimisation of the coupling matrix calculations. Section~\ref{sec:algorithm} derives the recursion relations that form the basis for the fast evaluation of $3j$ symbols. Section~\ref{sec:tests} benchmarks the algorithm using ACT DR6 survey window functions~\citep{naess2025atacamacosmologytelescopedr6} and compares performance with existing routines. We conclude in Sec.~\ref{sec:conclusion}, and two appendices provide further mathematical details.
\section{Pseudo-$C_{\ell}$ estimation}\label{sec:pseudo_cl}
In this section we outline the standard pseudo-$C_\ell$ algorithm for estimating
an unbiased power spectrum using mode-coupling matrices $K_{j_1 j_2}$ (see e.g. \cite{hivonMASTERCMBAnisotropy2002, efstathiouMythsTruthsConcerning2004}). For
notational consistency with the angular-momentum literature, we write $j$ in
place of $\ell$.
Since our focus is on accelerating the evaluation of Wigner--$3j$ symbols, here we
restrict attention to the TT and EE coupling matrices, which already contain the
two symbol configurations that appear in the full set of pseudo-$C_\ell$
estimators; fuller details are provided in Appendix \ref{sec:couple}.

At a given frequency, the expectation values of the pseudo-spectra satisfy
\begin{equation}
    \begin{aligned}
        \braket{\tilde{C}}^{T_i T_j} &= K^{T_i T_j} C^{TT},\\
        \braket{\tilde{C}}^{E_i E_j} &= K^{E_i E_j} C^{EE}
        + K^{E_i B_j} C^{BB},
    \end{aligned}
\label{eq:pseudo_cl}
\end{equation}
where $\tilde{C}^{X_i X_j}$ denotes the cross spectrum between two masked CMB maps $X_i$ and $X_j$, and $C^{TT}$, $C^{EE}$ and $C^{BB}$ denote the underlying temperature-, E-mode-, and B-mode-power spectra, respectively. The mode-coupling matrices take the form
\begin{align}
K^{T_i T_j}_{j_1 j_2}
&= \frac{(2j_2+1)}{4\pi}
   \sum_{j_3} (2j_3+1)\,\tilde{W}^{T_i T_j}_{j_3}
   \begin{pmatrix}
        j_1 & j_2 & j_3\\
        0   & 0   & 0
   \end{pmatrix}^2
\nonumber\\
&\equiv (2j_2+1)\,
\Xi_{TT}\!\left(j_1,j_2,\tilde{W}^{T_i T_j}\right),
\label{eq:K_TT}
\end{align}
\begin{align}
K^{E_i P_j}_{j_1 j_2}
&= \frac{(2j_2+1)}{4\pi}
   \sum_{j_3} (2j_3+1)\,\tilde{W}^{P_i P_j}_{j_3}
   \left(\frac{1 \pm (-1)^J}{2}\right)^2
\nonumber\\
&\quad\times
   \begin{pmatrix}
        j_1 & j_2 & j_3\\
        -2  & 2   & 0
   \end{pmatrix}^2
\nonumber\\
&\equiv (2j_2+1)\,
\Xi_{EP}\!\left(j_1,j_2,\tilde{W}^{P_i P_j}\right),
\label{eq:K_EE}
\end{align}
where $P_j\in\{E,B\}$ and $J\equiv j_1+j_2+j_3$. Here $\tilde{W}^{X_i Y_j}_\ell$ is the harmonic power spectrum of the survey window function, obtained from the auto- or cross-spectrum of the masks applied to the maps $X_i$ and $Y_j$.

The parity structure 
of these expressions is worth emphasising, in particular their dependence on whether $J$ is even or odd. The temperature
symbol
\[
\begin{pmatrix}
j_1 & j_2 & j_3\\
0 & 0 & 0
\end{pmatrix},
\]
which we may generically denote by $\mathcal{J}_{(0,0)}$, vanishes for odd $J$, so TT receives contributions only from even-parity
configurations.  In contrast, the polarization symbol
\[
\begin{pmatrix}
j_1 & j_2 & j_3\\
-2 & 2 & 0
\end{pmatrix},
\]
which we may generically denote by $\mathcal{J}_{(-2,2)}$, is non-zero for both even and odd $J$.  The relevant parity projection is
instead enforced by the factors $(1\pm (-1)^J)^2$ in
Eq.~\eqref{eq:K_EE}: the $+$ sign selects even $J$ (as in the EE matrix), while
the $-$ sign selects odd $J$ (as in the EB matrix).  Consequently, although the
$\mathcal{J}_{(-2,2)}$ symbol itself is non-zero for both parities, only its even-$J$ component enters
$K^{EE}$, while only the odd-$J$ component enters $K^{EB}$.  
These selection rules will be useful when deriving the optimized recurrence
relations in Section~\ref{sec:algorithm}, where we explicitly separate the
even- and odd-$J$ branches of the $\mathcal{J}_{(-2,2)}$ symbols and show how both can be
expressed in terms of the $\mathcal{J}_{(0,0)}$ symbols.

In summary, to compute all elements of the coupling matrices, we require Wigner--$3j$
symbols of the form
\begin{equation}
    \begin{pmatrix}
        j_1 & j_2 & j_3 \\
        0   & 0   & 0
    \end{pmatrix},
    \label{eq:3j_000}
\end{equation}
and
\begin{equation}
    \begin{pmatrix}
        j_1 & j_2 & j_3 \\
        -2  & 2   & 0
    \end{pmatrix}.
    \label{eq:3j_m220}
\end{equation}
In this work we use the calculation of the $K^{T_i T_j}$ and $K^{E_i E_j}$ mode-coupling matrices as representative benchmarks for the efficiency of our Wigner–$3j$ computations, noting that the same symbols also enter analytic pseudo-$C_\ell$ covariance calculations \citep{efstathiouMythsTruthsConcerning2004,brownCMBTemperaturePolarisation2005}.

As shown in \cite{efstathiouMythsTruthsConcerning2004}, the exact evaluation of
the covariance expression is computationally impractical at high multipoles, and one instead
considers an approximation motivated by a simplification that occurs when the CMB power spectrum does not vary over the band limit of the mask. In this limit, the pseudo-$C_\ell$ covariance reduces to a kernel
constructed from the power spectrum of the squared mask. 

Consequently, approximate covariance constructions in this regime involve exactly the
same classes of Wigner--$3j$ symbols as those appearing in the coupling matrices,
including both $(m_1,m_2,m_3)=(0,0,0)$ and, for polarization,
$(m_1,m_2,m_3)=(-2,2,0)$ configurations.
The fast evaluation techniques developed in this work therefore apply directly to the
efficient computation of such approximate pseudo-$C_\ell$ covariance terms.

We first simplify the computation of the $3j$
symbols with all $m$ indices equal to zero, i.e.\ the $\mathcal{J}_{(0,0)}$ above.
We then use recursion relations to express the symbols with $(m_1,m_2,m_3)=(-2,2,0)$, i.e.\ the $\mathcal{J}_{(-2,2)}$.  In deriving the latter, we also derive expressions for $3j$ symbols with $(m_1,m_2,m_3)=(-1,1,0)$, which we may denote $\mathcal{J}_{(-1,1)}$.

For the polarization coupling matrices that involve $3j$ symbols as in
Eq.~\eqref{eq:3j_m220}, our implementation first computes the equivalent
$(0,-2,2)$ configuration using the optimized recursion relations derived in
Section~\ref{sec:algorithm}, and then applies the permutation symmetry of the $3j$
symbols,
\begin{equation}
    \begin{pmatrix}
        j_1 & j_2 & j_3 \\
        m_1 & m_2 & m_3
    \end{pmatrix}
    =
    \begin{pmatrix}
        j_2 & j_3 & j_1 \\
        m_2 & m_3 & m_1
    \end{pmatrix},
    \label{eq:permutation}
\end{equation}
to obtain the required $(-2,2,0)$ configuration.

\section{Fast Computation Algorithm}
\label{sec:algorithm}
The mode-coupling matrices considered in this work depend on Wigner--$3j$ symbols
of the form
\[
\begin{pmatrix} j_1 & j_2 & j_3 \\ 0 & 0 & 0 \end{pmatrix},
\qquad
\begin{pmatrix} j_1 & j_2 & j_3 \\ -2 & 2 & 0 \end{pmatrix},
\]
which enter squared in the TT and EE coupling matrices (see
Eqs.~\ref{eq:K_TT} and~\ref{eq:K_EE}), and appear as products of the two symbols in the
TE case (see Eq.~\ref{eq:Sig_TE}).  

For fixed $(j_1,j_2)$ the coupling kernels require summation over $j_3$ in steps
of two, reflecting the constraint that $J=j_1+j_2+j_3$ must be even, for TT, TE, EE and BB, or odd, for EB. The
combinations
\begin{align}\label{eq:trianglecombos}
&J        = j_1 + j_2 + j_3,\nonumber \\
&J_{-++}  = -j_1 + j_2 + j_3,\nonumber \\
&J_{+-+}  =  j_1 - j_2 + j_3,\nonumber \\
&J_{++-}  =  j_1 + j_2 - j_3.
\end{align}
therefore change by $\pm 2$  when $j_3\mapsto j_3+2$. Since the allowed values of
$j_3$ enforce integer parity of these combinations, it is convenient to introduce
the halved variables
\[
p = J/2,\quad
p_1 = J_{-++}/2,\quad
p_2 = J_{+-+}/2,\quad
p_3 = J_{++-}/2,
\]
which change by $\pm1$ between successive terms in the sum. In coupling matrix calculations this
structure allows the summation over $j_3$ to be implemented as a simple recurrence,
with each new term obtained from the previous one using integer-index updates and a
small number of floating-point operations.

\subsection{Squared Symbol $\mathcal{J}_{(0,0)}$}
\label{sec:000}
Equations (3.7.4) and (3.7.5) of E57 state that a $3j$ symbol remains unchanged under even permutations of its columns while an odd permutation introduces a factor of $(-1)^J$. Additionally, from E57 Eq.\ (3.7.6), flipping the signs of all $m$-values also results in a factor of $(-1)^J$:  
\begin{equation}
  \begin{pmatrix}
    j_1 & j_2 &  j_3 \\ -m_1 & -m_2 & -m_3 
  \end{pmatrix}
  =
  (-1)^J
  \begin{pmatrix}
    j_1 & j_2 &  j_3 \\ m_1 & m_2 & m_3 
  \end{pmatrix}.
  \label{eq:mflip}
\end{equation}  
Thus for example such a symbol with $m_i=0$ may be non-zero only if $J$ is even.

By examining E57 Eq.\ (3.7.17), the square of the $m_i=0$ Wigner--$3j$ symbol can be expressed as 
\begin{equation}
\begin{pmatrix}
j_1 & j_2 & j_3 \\ 0 & 0 & 0
\end{pmatrix}^2
=
\frac{1}{J+1}\,
\frac{g(p_1)\, g(p_2)\, g(p_3)}{g(p)},
\label{eq:000_sq_main}
\end{equation}
where we have introduced the function
\begin{equation}
g(p)=\frac{(2p)!}{2^{\,2p}(p!)^2}.
\label{eq:gdef}
\end{equation}
The factors of $2^{2p}$ cancel in Eq.~\eqref{eq:000_sq_main}, but retaining them in
Eq.~\eqref{eq:gdef} keeps $g(p)$ slowly varying with $p$, which is advantageous for numerical work.
As a guide to its scaling, Stirling’s approximation gives $g(p)\sim (\pi p)^{-1/2}$ for large $p$.

To obtain the recurrence relation used in the algorithm, we rewrite the factorial
in Eq.~\eqref{eq:gdef} as
\[
(2p)! = 2p\,(2p-1)\,(2p-2)\cdots 1
      = 2^{\,2p}\,p!\,(p-\tfrac12)(p-\tfrac32)\cdots\tfrac12.
\]
Substituting this into Eq.~\eqref{eq:gdef} yields
\[
g(p)
= \frac{(p-\tfrac12)(p-\tfrac32)\cdots\tfrac12}{p!},
\]
from which the recurrence
\begin{equation}
g(p) = \frac{p-\tfrac12}{p}\, g(p-1)
\label{eq:g_rec}
\end{equation}
follows immediately. 

Starting from $g(0)=1$, Eq.~\eqref{eq:g_rec} is used to construct $\ln g(p)$ for
$p=1,\ldots,p_{\max}$ in linear time, after which both $g(p)$ and its reciprocal
$1/g(p)$ are obtained by exponentiation and stored, with
$p_{\max}\sim\mathcal{O}(\ell_{\max})$. The factors $(J+1)^{-1}$ required in
Eq.~\eqref{eq:000_sq_main} are likewise precomputed for the relevant range of
$J$. Once these tables are available, evaluation of the squared $(0,0,0)$ symbol
reduces to a small number of multiplications and cache-resident array lookups,
with no factorials or transcendental functions appearing in the main summation
loops.

\subsection{Squared Symbol $\mathcal{J}_{(-2,2)}$}
\label{sec:022}
The polarization kernels require the symbol
\[
\begin{pmatrix}
j_1 & j_2 & j_3 \\ -2 & 2 & 0
\end{pmatrix}=\mathcal{J}_{(-2,2)}.
\]
To obtain it, we first derive the intermediate expression for the
$(0,-2,2)$ configuration. Applying Eq.~(3.7.13) of E57 with $m_1=0$,
$m_2=-1$, $m_3=1$, and shifting $j_3\mapsto j_3+1$, gives
\begin{align}
  &
  \sqrt{(J+2)(J_{-++}+1)(J_{+-+}+1)J_{++-}}
  \begin{pmatrix}
      j_1 & j_2 & j_3+1 \\ 0 & -1 & 1
  \end{pmatrix}
  \nonumber \\
  &= 
  \lambda\,
  \begin{pmatrix}
      j_1 & j_2 & j_3 \\ 0 & 0 & 0
  \end{pmatrix}
  \nonumber \\
  &\quad+
  2\sqrt{j_3(j_3+2)}\,
  \begin{pmatrix}
      j_1 & j_2 & j_3 \\ 0 & -1 & 1
  \end{pmatrix}
  \nonumber \\
  &\quad-
  \eta\,
  \begin{pmatrix}
      j_1 & j_2 & j_3 \\ 0 & -2 & 2
  \end{pmatrix},
\label{eq:022_step}
\end{align}
with
\begin{equation}
\begin{aligned}
\lambda &= \sqrt{j_2(j_2+1)(j_3+1)(j_3+2)},\\
\eta &= \sqrt{(j_2-1)(j_2+2)(j_3-1)j_3}.
\end{aligned}
\end{equation}
The two $\mathcal{J}_{(-1,1)}$ symbols in Eq.~\eqref{eq:022_step} can be written entirely in
terms of $\mathcal{J}_{(0,0)}$ symbols with shifted arguments; these identities are derived
in Appendix~B. Substituting them into Eq.~\eqref{eq:022_step} yields
\begin{equation}
\begin{pmatrix}
j_1 & j_2 & j_3 \\ 0&-2&2
\end{pmatrix}
=
\eta^{-1}\!
\left[
\alpha
\begin{pmatrix}
j_1 & j_2 & j_3 \\ 0&0&0
\end{pmatrix}
+
\beta
\begin{pmatrix}
j_1 & j_2 & j_3\!+\!2 \\ 0&0&0
\end{pmatrix}
\right],
\label{eq:022_linear}
\end{equation}
with $\alpha$ and $\beta$ given in closed form in Appendix B. 
Since the coupling matrices involve only the square of this quantity, the
algorithm evaluates
\begin{equation}
\left(
\begin{matrix}
j_1 & j_2 & j_3 \\ 0&-2&2
\end{matrix}
\right)^2
=
\eta^{-2}
\left(
\alpha^2 X^2
+2\alpha\beta\,|XY|\,s(J)
+\beta^2 Y^2
\right),
\label{eq:022_sq}
\end{equation}
where $X$ and $Y$ denote the two $\mathcal{J}_{(0,0)}$ symbols, $|XY|=\sqrt{X^2Y^2}$, and
$s(J)$ is a parity-dependent sign factor. A final even permutation maps the
$(0,-2,2)$ symbol to the required $(-2,2,0)$ configuration, as in Eq.~\eqref{eq:permutation}.

\section{Speed tests of \ourcodename}
\label{sec:tests}

From this point onward we revert to the standard CMB notation and use $\ell$
(rather than $j$) for multipole indices; $\ell_{\max}$ denotes the maximum
multipole used in the coupling-matrix construction.
To assess performance, we benchmark the \ourcodename\space implementation  
against a reference implementation based on the standard Schulten--Gordon (S-G)
recursion, as used in the \texttt{pspy} package. The \texttt{pspy} module
\texttt{mcm\_compute} evaluates the mode-coupling matrices using the S-G recursion
and is representative of current pseudo-$C_\ell$ pipelines.

We benchmark using the ACT DR6 survey window map~\citep{naess2025atacamacosmologytelescopedr6}, which is the same for both temperature and polarization.
For each choice of $\ell_{\max}$, the mask auto--power spectrum $W_\ell$ is computed
directly from this map using \texttt{pixell}~\citep{2021ascl.soft02003N} by evaluating
the spherical-harmonic coefficients of the window up to $2\ell_{\max}$ and forming the
auto-spectrum
$W_\ell = \frac{1}{2\ell+1}\sum_m |w_{\ell m}|^2$.
The factor of $2\ell_{\max}$ reflects the standard pseudo-$C_\ell$ formalism, which
requires $W_L$ for $L \le 2\ell_{\max}$ in the coupling-matrix construction.
Timing tests are performed for ten values of $\ell_{\max}$ linearly spaced between
$500$ and $10{,}000$.
For each $\ell_{\max}$, the resulting $W_\ell$ is passed identically to both
\ourcodename\space and the S-G-based reference implementation, and the
wall-clock runtime of each call is measured.
Each test is repeated five times, and we report the mean execution time and standard
deviation.
All CPU benchmarks are performed on an 8-core Apple M3 processor.

\begin{figure*}
    \centering
    \includegraphics[width=0.85\linewidth]{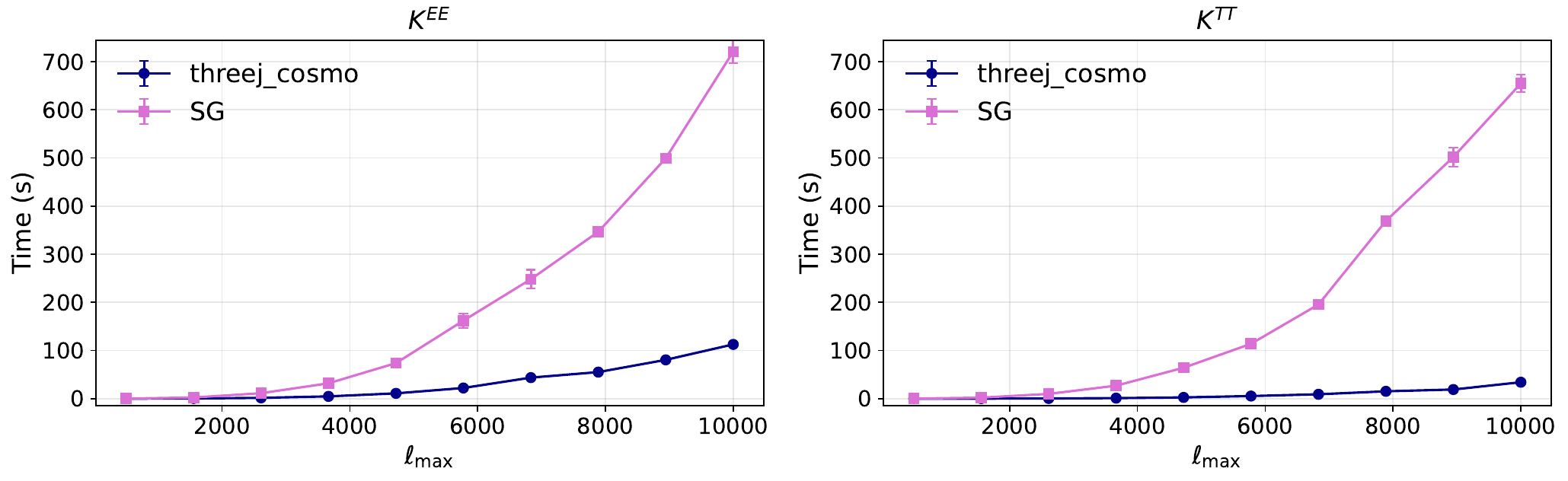}
    \caption{CPU time for the calculation of the mode-coupling matrices
    $K^{EE}$ (left) and $K^{TT}$ (right) as a function of $\ell_{\max}$,
    comparing the \ourcodename\space implementation with the reference
    Schulten--Gordon (S-G) algorithm.
    Tests use the ACT DR6 mask and were run on an
    8-core Apple M3 CPU. Error bars show the standard deviation over 5 runs
    per $\ell_{\max}$. For TT, \ourcodename\space achieves speedups of
    $\sim 20$–$25\times$ for $\ell_{\max}\gtrsim 2000$
    (e.g.\ $25.3\times$ at $\ell_{\max}=4722$), while for EE the speedup is
    typically $\sim 6$–$7\times$ (e.g.\ $6.7\times$ at $\ell_{\max}=4722$).}
    \label{fig:cpu_time_act}
\end{figure*}
\begin{figure}
    \centering
    \includegraphics[width=\linewidth]{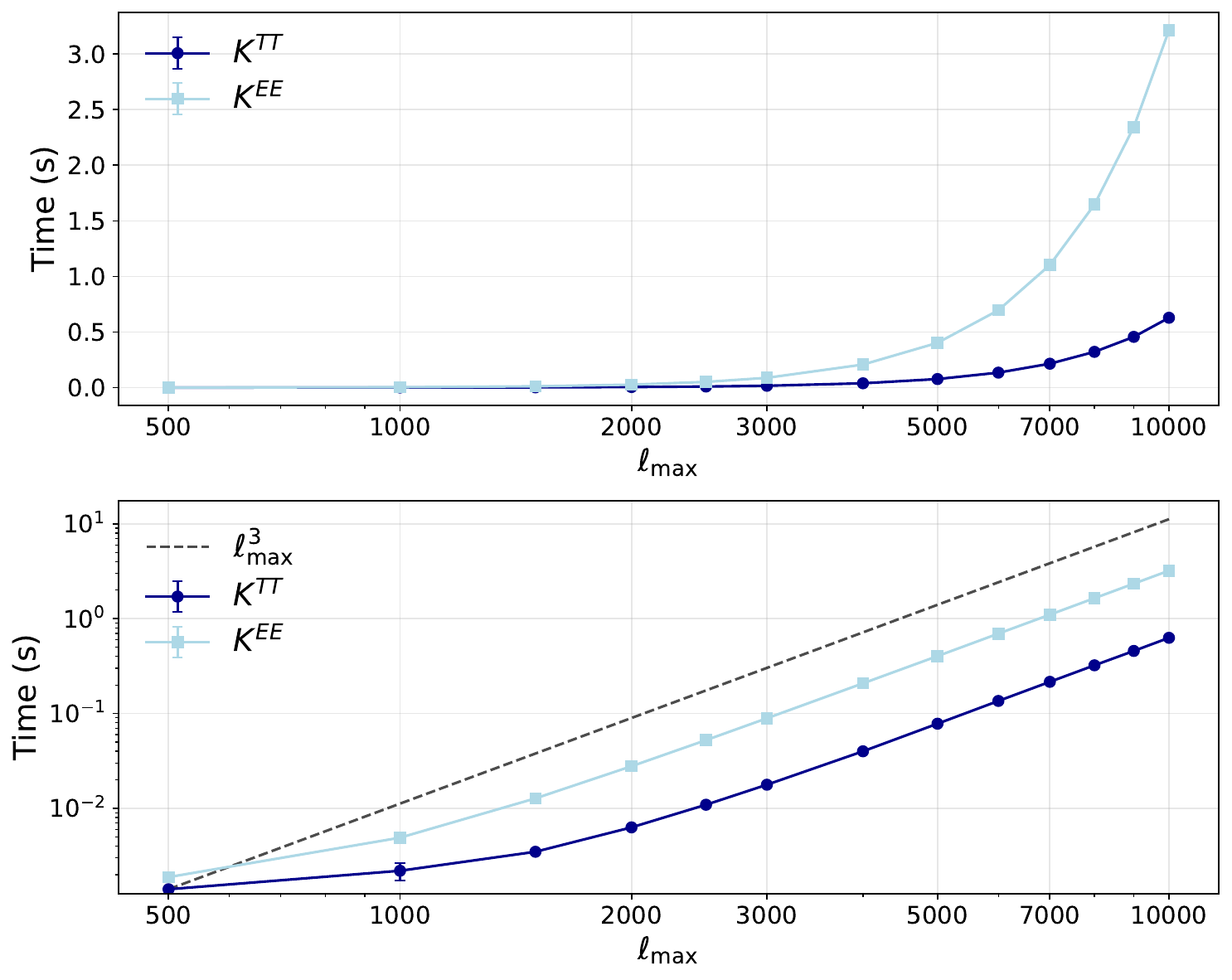}
\caption{Kernel execution time as a function of multipole moment $\ell_{\mathrm{max}}$
    for the GPU-accelerated \ourcodename\ code computing $K^{TT}$ and $K^{EE}$
    coupling matrices. Benchmarks were performed on an NVIDIA A100 GPU (40\,GB)
    using OpenMP target offloading with linearized triangular iteration for full
    parallelization. Times shown are kernel execution only, excluding data transfer
    overhead (${\sim}0.4$\,s per invocation). Error bars show the standard deviation 
    over 5 runs.
    \emph{Top:} Linear scale showing $K^{TT}$ computation completes in 0.63\,s
    and $K^{EE}$ in 3.2\,s at $\ell_{\max} = 10^4$.
    \emph{Bottom:} Log-log scale with $\ell_{\max}^3$ reference line (dashed),
    confirming the expected cubic scaling.}
    \label{fig:gpu_time}
\end{figure}

\subsection{CPU performance}\label{sec:CPU}
As seen in Figure~\ref{fig:cpu_time_act}, the \ourcodename\space implementation
exhibits the expected cubic scaling,
$T \propto \ell_{\max}^3$, but with a substantially reduced prefactor compared
to the reference S-G recursion.
This reduction leads to large and consistent performance gains across the full
range of multipoles tested.

For TT coupling matrices, the speedup increases rapidly with $\ell_{\max}$,
reaching factors of $20$–$25\times$ for $\ell_{\max}\gtrsim2000$ and remaining
at least $19\times$ faster up to $\ell_{\max}=10^4$.
For example, at $\ell_{\max}=4722$ the S-G-based runtime is $64.2$\,s, while
\ourcodename\space completes the same calculation in $2.54$\,s.
At $\ell_{\max}=10^4$, the TT computation requires $\sim34$\,s with
\ourcodename, compared with over ten minutes for the S-G implementation.

The EE coupling matrices follow the same overall scaling behaviour but with a
smaller relative speedup.
At low multipoles the improvement is modest ($\sim1.7\times$ at
$\ell_{\max}=500$), stabilising to $6$–$7\times$ for
$\ell_{\max}\gtrsim1500$.
This difference reflects the additional algebraic operations
associated with the $\mathcal{J}_{(-2,2)}$ configuration.

The \ourcodename\space implementation parallelises the outer loop over $\ell_1$
using OpenMP with dynamic scheduling.
All intermediate quantities are thread-private, and the coupling matrix is
stored in row-major order to ensure contiguous memory access.
After computing the upper triangle, the symmetry
$K_{\ell_1 \ell_2}=K_{\ell_2 \ell_1}$ is used to fill in the lower triangle.
This strategy provides good load balance across the triangular iteration space
and avoids redundant computation.

\subsection{GPU acceleration}

We additionally benchmark a GPU implementation based on OpenMP target
offloading, compiled with the NVIDIA HPC SDK and executed on an NVIDIA A100
GPU (40\,GB).
The GPU kernel reuses the same algorithm as the CPU
implementation: each $(\ell_1,\ell_2)$ element is computed independently and
maps naturally onto fine-grained parallelism.

Figure~\ref{fig:gpu_time} shows the GPU performance when measuring kernel
execution time only, excluding data transfer and kernel-launch overheads.
Under this like-for-like comparison, the GPU implementation outperforms the S-G implementation across the entire multipole range
tested, including the lowest values of $\ell_{\max}$.
At $\ell_{\max}=10^4$, the GPU kernel time is $0.63\,\mathrm{s}$ for TT and
$3.2\,\mathrm{s}$ for EE. This corresponds to speedups of
$54\times$ for TT and $35\times$ for EE when compared to the optimized
\ourcodename\space CPU implementation, and exceeds $1000\times$ for TT and
$200\times$ for EE when compared directly to the S-G baseline.

The measured GPU kernel times scale as $t \propto \ell_{\max}^3$ for both TT and
EE, in close agreement with the theoretical arithmetic complexity of the
algorithm.

\subsection{Band-limited covariance matrices}
As discussed earlier, approximate pseudo-$C_\ell$ covariance expressions for
apodized or otherwise band-limited masks lead to coupling-matrix-like kernels
in which the sum over the intermediate multipole $\ell_3$ is effectively
truncated at a characteristic scale $L_{\rm apo}$ set by the support of the
survey window function in harmonic space.
In this regime, the incremental algorithm can exploit the reduced coupling
width by terminating the internal $\ell_3$ summation at a cutoff $L_{\rm cut}$.

To isolate the performance impact of this effect, we benchmark a modified
implementation in which the window-function power spectrum
$\tilde{W}_{\ell_3}$ (defined as the auto-spectrum of the real-space mask)
satisfies $\tilde{W}_{\ell_3}=0$ for $\ell_3 > L_{\rm cut}$, so that truncating the
summation introduces no approximation.
This benchmarking is performed on the CPU setup described in
Section~\ref{sec:CPU}.
Numerical agreement with both the uncapped \ourcodename\space implementation
and the S-G-based reference is verified at the $10^{-10}$–$10^{-14}$ level across
all tested multipoles.

In this band-limited setting, speedups relative to the S-G reference become very large, and depend on both $\ell_{\max}$ and $L_{\rm cut}$. At fixed $L_{\rm cut}$, the speedup grows with $\ell_{\max}$: for TT coupling at $L_{\rm cut}=64$, improvements increase from $\sim28\times$ at $\ell_{\max}=1500$ to $\sim115\times$ at $\ell_{\max}=5000$. At fixed $\ell_{\max}$, the speedup decreases as $L_{\rm cut}$ approaches $\ell_{\max}$, since the effective summation range widens. For example at $\ell_{\max}=3000$, TT speedups range from $\sim60\times$ at $L_{\rm cut}=16$ down to $\sim21\times$ at $L_{\rm cut}=512$. Polarization coupling matrices show comparable or larger gains across the same parameter space, reaching $\sim480\times$ for EE at $\ell_{\max}=5000$ with small $L_{\rm cut}$. By contrast, the runtime of the S-G implementation is essentially independent of $L_{\rm cut}$, since it evaluates the full $\ell_3$ range for each matrix element.

\subsection{Discussion}
The benchmarks in Fig.~\ref{fig:cpu_time_act} show that the new
algorithm delivers substantial and robust speedups over the standard
S-G-based Fortran implementation.
On the CPU, the runtime follows the expected $\ell_{\max}^3$ scaling but with a
significantly reduced prefactor, leading to speedups of
$20$–$25\times$ for TT and $6$–$7\times$ for EE at high multipoles.
Across all tested configurations, numerical agreement with the Fortran
reference is maintained at the level of $10^{-12}$.

The GPU results in Fig.~\ref{fig:gpu_time} illustrate a complementary
performance regime.
While overheads are significant at low $\ell_{\max}$, the GPU becomes increasingly
effective for large matrices, ultimately providing an additional order-of-magnitude
acceleration relative to the CPU for TT and a factor of several for
EE.
This behaviour is consistent with the parallel evaluation of independent
$(j_1,j_2)$ blocks and the arithmetic intensity of the kernel at
high multipoles.

In addition to full matrix construction, the structure of the algorithm is particularly well matched to band-limited coupling calculations. In this case, the incremental formulation allows unnecessary recursion steps to be bypassed, yielding further reductions in computational cost without modifying the core kernel or sacrificing accuracy. This flexibility is relevant
for applications involving apodized masks or approximate covariance terms,
where coupling support is limited.

\section{Conclusion}\label{sec:conclusion}
We have presented an efficient method for evaluating the restricted set of Wigner--$3j$ symbol combinations that arise in CMB pseudo-$C_\ell$ mode-coupling calculations.
By expressing both temperature and polarization couplings in terms of a small number
of squared $\mathcal{J}_{(0,0)}$ symbols, and obtaining more complex configurations
algebraically from these base quantities, the method avoids redundant calculations.

The resulting implementation retains the formal $\mathcal{O}(\ell_{\max}^3)$ scaling
of standard approaches but with a reduced computational cost, while
maintaining numerical accuracy at the level required for precision CMB analyses.
The same kernel is readily parallelised on both CPUs and GPUs, enabling
efficient execution across a range of architectures.

The formulation is suited to scenarios in which only a
restricted set of coupling modes is required, such as band-limited coupling matrices
and analytic covariance constructions.
As a result, \ourcodename{} provides a lightweight, accurate, and scalable backend for
pseudo-$C_\ell$ pipelines, with direct relevance for current and upcoming CMB
experiments.

Our approach could be extended beyond the $\mathcal{J}_{(0,0)}$,\space  $\mathcal{J}_{(-1,1)}$ and \space $\mathcal{J}_{(-2,2)}$ symbols. For example in the analysis of the stochastic gravitational wave background (see e.g. \cite{PhysRevD.110.063547}) the symbols with magnetic numbers $(1,1,-2)$ and $(2,2,-4)$ also appear. It would be interesting to see if similar speed-ups are achievable for these symbols also.

\section*{Acknowledgements}

We thank Anthony Challinor and George Efstathiou for useful discussions and Roger de Belsunce for helpful comments on a draft of this paper.
This work was performed using resources provided by the Cambridge Service for Data Driven Discovery (CSD3) operated by the University of Cambridge Research Computing Service (\href{https://www.csd3.cam.ac.uk}{www.csd3.cam.ac.uk}), provided by Dell EMC and Intel using Tier-2 funding from the Engineering and Physical Sciences Research Council (capital grant EP/T022159/1), and DiRAC funding from the Science and Technology Facilities Council (\href{https://www.dirac.ac.uk}{www.dirac.ac.uk})
funded by BEIS capital funding via STFC Capital Grants
ST/P002307/1 and ST/R002452/1 and STFC Operations Grant ST/R00689X/1. The authors acknowledge funding from EPSRC ExCALIBUR H\&SE grant EP/Y028082/1 and
STFC Astronomy Consolidated Grant ST/X001113/1. G.K. is supported by the STFC CDT (Data Intensive Science) grant ST/W006812/1.

\section*{Data Availability}

The code developed for this work, \ourcodename, is publicly available at \url{https://github.com/gkiddier22/threej_cosmo}. No new observational data were generated or analysed in this study.

\bibliography{references}

\clearpage
\onecolumn
\appendix
\section{Coupling Matrices}
\label{sec:couple}

The pseudo-$C_\ell$ formalism expresses the mode coupling induced by a survey
window function through a set of geometry-dependent kernels $\Xi_{XY}$, where
$X,Y\in\{T,E,B\}$. Each coupling matrix can be written in the generic form
\[
K^{X_i Y_j}_{\ell_1 \ell_2}
= (2\ell_2+1)\,\Xi_{XY}(\ell_1,\ell_2,\tilde{W}^{X_i Y_j}),
\]
where $\tilde{W}^{X_i Y_j}_\ell$ denotes the harmonic power spectrum of the
survey window function for the map pair $(X_i,Y_j)$.
The explicit form of $\Xi_{XY}$ is determined by combinations of Wigner--$3j$
symbols and is given below for each case.

All expressions in this section use the shorthand
\[
L \equiv \ell_1 + \ell_2 + \ell_3,
\]
where $\ell_3$ denotes the intermediate multipole summed over in the
coupling-matrix construction.

\paragraph*{Temperature–temperature (TT).}
Only the symbol with $m_i=0$ appears:
\begin{align}
\Xi_{TT}(\ell_1,\ell_2,\tilde{W})
&= \frac{1}{4\pi}
\sum_{\ell_3} (2\ell_3+1)\,
\tilde{W}_{\ell_3}\,
\begin{pmatrix}
\ell_1 & \ell_2 & \ell_3\\
0 & 0 & 0
\end{pmatrix}^{\!2}.
\end{align}
\paragraph*{Temperature–polarization (TE).}
The TE coupling involves the product of the $m_i=0$ and
$m_1=-2,m_2=2,m_3=0$ Wigner--$3j$ symbols and is non-zero only for even $L$:
\begin{align}
\Xi_{TE}(\ell_1,\ell_2,\tilde{W})
&= \frac{1}{4\pi}
\sum_{\ell_3} (2\ell_3+1)\,
\tilde{W}_{\ell_3}\,
\frac{1+(-1)^L}{2}\,
\begin{pmatrix}
\ell_1 & \ell_2 & \ell_3\\
0 & 0 & 0
\end{pmatrix}
\begin{pmatrix}
\ell_1 & \ell_2 & \ell_3\\
-2 & 2 & 0
\end{pmatrix}.
\label{eq:Sig_TE}
\end{align}

\paragraph*{E–E polarization (EE).}
This case involves only the $m_1=-2,m_2=2,m_3=0$ symbols and is likewise restricted
to even $L$:
\begin{align}
\Xi_{EE}(\ell_1,\ell_2,\tilde{W})
&= \frac{1}{4\pi}
\sum_{\ell_3} (2\ell_3+1)\,
\tilde{W}_{\ell_3}\,
\left[\frac{1+(-1)^L}{2}\right]^2
\begin{pmatrix}
\ell_1 & \ell_2 & \ell_3\\
-2 & 2 & 0
\end{pmatrix}^{\!2}.
\end{align}

\paragraph*{E–B and B–E (EB/BE).}
These couplings vanish for even $L$ and are non-zero only for odd parity:
\begin{align}
\Xi_{EB}(\ell_1,\ell_2,\tilde{W})
&= \frac{1}{4\pi}
\sum_{\ell_3} (2\ell_3+1)\,
\tilde{W}_{\ell_3}\,
\left[\frac{1-(-1)^L}{2}\right]^2
\begin{pmatrix}
\ell_1 & \ell_2 & \ell_3\\
-2 & 2 & 0
\end{pmatrix}^{\!2}.
\end{align}
The same expression holds for $\Xi_{BE}$.

\paragraph*{B–B polarization (BB).}
The BB coupling kernel has the same structure and parity selection as the EE
case, differing only in the underlying power spectrum entering the
pseudo-$C_\ell$ relation.

\section{$3\MakeLowercase{j}$ Symbols}
\subsection{Obtaining $\mathcal{J}_{(-1,1)}$ symbols}
Using Eqs. (3.7.13) and (3.7.15) of E57, and Eq.~(\ref{eq:mflip}), gives for odd $J$:
\begin{align}
  \wigthreej{j_1}{j_2}{j_3}{0}{-1}{1}
  &= -\frac{1}{2} \left[\frac{(J+2)(J_{-++}+1)(J_{+-+}+1)(J_{++-})}{j_2 (j_2+1) j_3 (j_3+1)}\right]^{1/2} \times\wigthreej{j_1}{j_2}{j_3+1}{0}{0}{0}. \label{eq:joneodd}
\end{align}
For even $J$, we apply Eq.\ (3.7.12) of E57 twice. First, for $(m_1, m_2, m_3)=(0,-\frac{1}{2},\frac{1}{2})$ with $j_2 \mapsto j_2+\frac{1}{2}$ and $j_3 \mapsto j_3+\frac{1}{2}$:
\begin{equation}
  \sqrt{(J+2)(J_{-++}+1)} \wigthreej{j_1}{j_2+\frac{1}{2}}{j_3+\frac{1}{2}}{0}{-\frac{1}{2}}{\frac{1}{2}}
  = \sqrt{j_2  j_3} \wigthreej{j_1}{j_2}{j_3}{0}{-1}{1} -\sqrt{(j_2+1) (j_3+ 1)}\wigthreej{j_1}{j_2}{j_3}{0}{0}{0}. \label{eq:jeven}
\end{equation}
Applying Eq.\ (3.7.12) of E57 again with $m=0$ and shifting $j_2 \mapsto j_2+1$, $j_3 \mapsto j_3+1$:
\begin{equation}
   \wigthreej{j_1}{j_2+\frac{1}{2}}{j_3+\frac{1}{2}}{0}{-\frac{1}{2}}{\frac{1}{2}}
   = \frac{1}{2}\left[\frac{(J+3)(J_{-++}+2)}{(j_2+1)( j_3+1)}\right]^{1/2}\wigthreej{j_1}{j_2+1}{j_3+1}{0}{0}{0}. \label{eq:jhalf}
\end{equation}
Substituting Eq.~(\ref{eq:jhalf}) into Eq.~(\ref{eq:jeven}), we obtain
\begin{align}
   \wigthreej{j_1}{j_2}{j_3}{0}{-1}{1}  
   &= \left[ \frac{(j_2+1)(j_3+1)}{j_2 j_3} \right]^{1/2} \wigthreej{j_1}{j_2}{j_3}{0}{0}{0}
   + \frac{1}{2} \left[\frac{(J+2)(J+3)(J_{-++}+1)(J_{-++}+2)}{j_2 (j_2+1) j_3 (j_3+1)}\right]^{1/2}\times\wigthreej{j_1}{j_2+1}{j_3+1}{0}{0}{0}. \label{eq:joneeven}
\end{align}

\subsection{Further Expressions for $\mathcal{J}_{(-1,1)}$ symbols}
From Eq.~(\ref{eq:022_step}), we need $m_1=0,m_2=-1,m_3=1$ symbols for both $(j_1, j_2, j_3)$ and $(j_1, j_2, j_3+1)$.
For odd $J$, Eq.~(\ref{eq:joneodd}) provides $\wigthreej{j_1}{j_2}{j_3}{0}{-1}{1}$, while for $\wigthreej{j_1}{j_2}{j_3+1}{0}{-1}{1}$, we shift $j_3 \mapsto j_3+1$ in Eq.~(\ref{eq:joneeven}):
\begin{align}
   \wigthreej{j_1}{j_2}{j_3+1}{0}{-1}{1}
   &= \left[ \frac{(j_2+1)(j_3+2)}{j_2( j_3+1)} \right]^{1/2} \wigthreej{j_1}{j_2}{j_3+1}{0}{0}{0} + \frac{1}{2} \left[\frac{(J+3)(J+4)(J_{-++}+2)(J_{-++}+3)}{j_2 (j_2+1) (j_3+1) (j_3+2)}\right]^{1/2} \times\wigthreej{j_1}{j_2+1}{j_3+2}{0}{0}{0}. \label{eq:joneevenshifted}
\end{align}
If $J$ is even, we can directly use Eq.~(\ref{eq:joneeven}) to obtain $\wigthreej{j_1}{j_2}{j_3}{0}{-1}{1}$. However, for $\wigthreej{j_1}{j_2}{j_3+1}{0}{-1}{1}$, we must use Eq.~(\ref{eq:joneodd}), shifting $j_3$ to $j_3+1$:
\begin{align}
   \wigthreej{j_1}{j_2}{j_3+1}{0}{-1}{1}
   &= -\frac{1}{2} \left[\frac{(J+3)(J_{-++}+2)(J_{+-+}+2)(J_{++-}-1)}
   {j_2 (j_2+1) (j_3+1) (j_3+2)}\right]^{1/2}\times \wigthreej{j_1}{j_2}{j_3+2}{0}{0}{0}.
   \label{eq:joneoddshifted}
\end{align}

\subsection{Closed-form reduction of $\mathcal{J}_{(-2,2)}$}
\label{app:022_closed_form}

We derive the two-term closed form for the $\mathcal{J}_{(-2,2)}$ configuration used in
the implementation. We define
\begin{align}
\lambda &\equiv \sqrt{j_2(j_2+1)(j_3+1)(j_3+2)}, \nonumber \\
\eta &\equiv \sqrt{(j_2-1)(j_2+2)(j_3-1)j_3},
\end{align}
and use the standard triangle combinations as in Eq.~\eqref{eq:trianglecombos}.

Starting from Eq.~\eqref{eq:022_step}, we rewrite it as
\begin{align}
\eta\,
\wigthreej{j_1}{j_2}{j_3}{0}{-2}{2}
&=
\lambda\,\wigthreej{j_1}{j_2}{j_3}{0}{0}{0}
+2\sqrt{j_3(j_3+2)}\,\wigthreej{j_1}{j_2}{j_3}{0}{-1}{1} -\Lambda\,
\wigthreej{j_1}{j_2}{j_3+1}{0}{-1}{1},
\label{eq:022_rearranged}
\end{align}
where
\begin{equation}
\Lambda \equiv \sqrt{(J+2)(J_{-++}+1)(J_{+-+}+1)J_{++-}}.
\end{equation}

For even $J$, the symbol $\wigthreej{j_1}{j_2}{j_3}{0}{-1}{1}$ reduces to a single
$(0,0,0)$ symbol,
\begin{align}
\wigthreej{j_1}{j_2}{j_3}{0}{-1}{1}
&=
\sqrt{\frac{(j_2+1)(j_3+1)}{j_2 j_3}}\,
\left(1-\frac{(J+2)(J_{-++}+1)}{2(j_2+1)(j_3+1)}\right) \times \wigthreej{j_1}{j_2}{j_3}{0}{0}{0},
\label{eq:jone_even_singleterm}
\end{align}
while the shifted symbol $\wigthreej{j_1}{j_2}{j_3+1}{0}{-1}{1}$ has odd total angular
momentum and is therefore given by
\begin{align}
\wigthreej{j_1}{j_2}{j_3+1}{0}{-1}{1}
&=
-\frac12
\left[
\frac{(J+3)(J_{-++}+2)(J_{+-+}+2)(J_{++-}-1)}
{j_2(j_2+1)(j_3+1)(j_3+2)}
\right]^{1/2} \times \wigthreej{j_1}{j_2}{j_3+2}{0}{0}{0}.
\label{eq:jone_odd_for_shift}
\end{align}

Substituting Eq.~\eqref{eq:jone_even_singleterm} into the second term of
Eq.~\eqref{eq:022_rearranged} gives
\begin{equation}
    2\sqrt{j_3(j_3+2)}\,
    \wigthreej{j_1}{j_2}{j_3}{0}{-1}{1}
    =
    \frac{2\lambda}{j_2}
    \left(1-\frac{(J+2)(J_{-++}+1)}{2(j_2+1)(j_3+1)}\right)
    \wigthreej{j_1}{j_2}{j_3}{0}{0}{0},
    \label{eq:term2_simplified}
\end{equation}
where we used $\lambda^2=j_2(j_2+1)(j_3+1)(j_3+2)$.

Similarly, substituting Eq.~\eqref{eq:jone_odd_for_shift} into the third term of
Eq.~\eqref{eq:022_rearranged} yields
\begin{align}
-\Lambda\,
\wigthreej{j_1}{j_2}{j_3+1}{0}{-1}{1}
&=
\frac{1}{2\lambda}
\bigl[
(J+2)(J_{-++}+1)(J_{+-+}+1)J_{++-} \times (J+3)(J_{-++}+2)(J_{+-+}+2)(J_{++-}-1)
\bigr]^{1/2} \times \wigthreej{j_1}{j_2}{j_3+2}{0}{0}{0}.
\label{eq:term3_simplified}
\end{align}

Collecting terms and dividing by $\eta$, we obtain the final two-term reduction
\begin{equation}
\wigthreej{j_1}{j_2}{j_3}{0}{-2}{2}
=
\eta^{-1}
\left[
\alpha\,
\wigthreej{j_1}{j_2}{j_3}{0}{0}{0}
+
\beta\,
\wigthreej{j_1}{j_2}{j_3+2}{0}{0}{0}
\right],
\label{eq:022_linear_closedform}
\end{equation}
with coefficients
\begin{align}
\alpha
&=
\lambda
+\frac{2\lambda}{j_2}
\left(1-\frac{(J+2)(J_{-++}+1)}{2(j_2+1)(j_3+1)}\right),
\\[3pt]
\beta
&=
\frac{1}{2\lambda}
\bigl[
(J+2)(J_{-++}+1)(J_{+-+}+1)J_{++-} \times (J+3)(J_{-++}+2)(J_{+-+}+2)(J_{++-}-1)
\bigr]^{1/2}.
\end{align}

Equation~\eqref{eq:022_linear_closedform} is used in the \ourcodename \space
implementation: the $\mathcal{J}_{(-2,2)}$ symbol is evaluated using two
$\mathcal{J}_{(0,0)}$ symbols at $(j_1,j_2,j_3)$ and $(j_1,j_2,j_3+2)$, together with
the standard parity phase relating the two $(0,0,0)$ symbols.
\subsection{Possible simplification for even $J$ for $\mathcal{J}_{(-1,1)}$ symbols}
As written, Eq.~(\ref{eq:joneeven}) requires the evaluation of two $\mathcal{J}_{(0,0)}$ symbols. However, we can use Eqs.~(\ref{eq:000_sq_main}) and (\ref{eq:g_rec}) to write the square of one in terms of the square of the other (and then use Eq.\ (3.7.17) of E57 to restore the sign). We find
\begin{align}
    \wigthreej{j_1}{j_2+1}{j_3+1}{0}{0}{0}
    &= -\left[\frac{(J+2)(J_{-++}+1)}{(J+3)(J_{-++}+2)} \right]^{1/2}\times\wigthreej{j_1}{j_2}{j_3}{0}{0}{0}.
\end{align}
Using this, Eq.~(\ref{eq:joneeven}) gives
\begin{align}
    \wigthreej{j_1}{j_2}{j_3}{0}{-1}{1}
    &= \left[ \frac{(j_2+1)(j_3+1)}{j_2 j_3} \right]^{1/2}
    \left(1 - \frac{(J+2)(J_{-++}+1)}{2(j_2+1)(j_3+1)} \right)\times \wigthreej{j_1}{j_2}{j_3}{0}{0}{0}. \label{eq:joneevensimp}
\end{align}
\label{lastpage}
\end{document}